# LINKING THE FOUNDATIONS OF PHYSICS AND MATHEMATICS

D.J. BENDANIEL
Cornell University
Ithaca NY, 14853, USA

**Abstract.** The concept of definability of physical fields within a set-theoretical foundation is introduced. We propose an axiomatic set theory and show the Schrödinger equation and, more generally, a nonlinear sigma model come naturally out of the mathematics of this theory when a physical null postulate is added. Space-time is relational in this theory and quantization of the field proves to be equivalent to definability. Some additional examples of applicability to physics are given.

We look to provide a deep connection between physics and mathematics by requiring that physical fields be definable within a set-theoretical foundation. The usual set-theoretical foundation of mathematics is called Zermelo-Fraenkel (ZF). In ZF, a set U of finite integers is definable if and only if there exists a formula $\Phi_U(n)$ from which we can unequivocally determine whether a given finite integer n is a member of U or not. That is, when a set of finite integers is not definable, then there will be at least one finite integer for which it is impossible to determine whether it is in the set or not. Other sets are definable in a theory if and only if they can be mirrored by a definable set of finite integers. Most sets of finite integers in ZF are not definable. Moreover, the set of definable sets of finite integers is itself not definable in ZF. [1]

A physical field in a finite region of space is definable in a set-theoretical foundation if and only if the set of possible distributions of the field's energy among its eigenstates can be mirrored by a definable set of finite integers in the theory. This concept of definability is physically meaningful because, were there a field whose set of energy distributions among eigenstates corresponded to an undefinable set of finite integers, that field would have at least one energy distribution whose presence or absence is cannot be determined, so the field could not be verifiable or falsifiable. Therefore, our task is to find a foundation in which we are able to specify completely the definable sets of finite integers but which also contains mathematics rich enough for the physical fields corresponding to these sets.

The definable sets of finite integers cannot be specified completely in ZF because that theory contains infinite sets whose definability is undecidable. So we must start with a sub-theory of ZF containing no infinite sets of finite integers. Then all sets of finite integers are *ipso facto* definable. This means, of course, that the set of all finite integers, called ω, cannot exist in that sub-theory. The set ω exists in ZF essentially in consequence of two axioms: an axiom of infinity and an axiom schema of subsets. Thus, we must delete one or the other of these axioms. If we delete the axiom of infinity we will then have no need for the axiom schema of subsets either, since all sets are finite. However that theory is not rich enough to obtain continuous functions of a real variable, so it is not useful for physical fields. The task reduces to whether or not, starting with ZF without the axiom schema of subsets, we can build with additional axioms a theory rich enough for the mathematics necessary for physical fields.

In the appendix we show eight axioms. The first seven are the axioms of ZF except that the axiom schema of replacement has been modified. The usual replacement axiom (AR) asserts that, for any functional relation, if the domain is a set, then the range is a set. That axiom actually combines two independent axioms: the axiom schema of subsets, which we wish to delete, and an axiom schema of bijective replacement (ABR), which refers only to a one-to-one functional relation. We can thus delete the axiom schema of subsets from ZF by substituting ABR for AR, forming the sub-theory ZF–AR+ABR.

We shall first see how ZF–AR+ABR differs importantly from ZF by looking at the axiom of infinity. The axiom of infinity asserts the existence of at least one set ω*. There are actually many such sets. In ZF, we can obtain the minimal ω*, a set usually called ω, by using the axiom schema of subsets to provide the intersection of all the sets created by the axiom of infinity. However, without the axiom schema of subsets the set ω cannot be obtained so that all theorems of ZF–AR+ABR must hold for any ω*. Any member of any ω* is an "integer". A "finite integer" is a member of every ω*. We shall use *i, j, k, ℓ, m* or *n* to denote finite integers. An "infinite integer" is a member of ω* that is not a finite integer.

We can now adjoin to ZF–AR+ABR an axiom asserting that all subsets of any ω* are constructible. By constructible sets we mean sets that are generated sequentially by some procedure, one after the other, so that the procedure well-orders the sets. Gödel has shown that an axiom asserting that all sets are constructible can be consistently added to ZF, giving a theory called ZFC[+].[2] It has also been shown that no more than countably many subsets of ω* can be proven to exist in ZFC[+].[3] This result will, of course, hold for its sub-theory ZFC[+]–AR+ABR. Therefore we can adjoin to ZF–AR+ABR another axiom asserting that the subsets of any ω* are constructible and that there are no more than countably many such subsets. We call these eight axioms theory T.

Cantor's proof or its equivalent cannot be shown in T [4] thus no uncountably infinite sets exist. Since all sets are countable, the general continuum hypothesis will hold. Furthermore, in T all sets of finite integers are finite, so that any infinite set of integers must contain infinitely many infinite integers. Moreover, we cannot show the induction theorem, so that not even all the countable sets that exist in ZF are contained in T. For example, we cannot sum infinite series, whereas in ZF infinite series play an important role in the development of mathematics. Nevertheless, our axiom of constructibility will provide an alternate route for obtaining functions of a real variable, but only with rather special properties.

We first show that T actually contains a real line. Recall the definition of "rational numbers" as the set of ratios, usually called **Q**, of any two members of the set ω. In T, we can likewise, using the axiom of unions, establish for any ω* the set of ratios of any two of its integers, finite or infinite. This will be an "enlargement" of the rational numbers and we shall call this enlargement **Q***. Two members of **Q*** are called "identical" if their ratio is 1. We employ the symbol "≡" for "is identical to". An "infinitesimal" is a member of **Q*** "equal" to 0, i.e., letting $y$ signify the member and employing the symbol "=" to signify equality, $y = 0 \leftrightarrow \forall k[y < 1/k]$. The reciprocal of an infinitesimal is "infinite". Any member of **Q*** that is not an infinitesimal and not infinite is "finite", $[y \neq 0 \wedge 1/y \neq 0] \leftrightarrow \exists k[1/k < y < k]$.

We apply this concept of equality to the interval between two finite members of $\mathbf{Q}^*$; two finite members are either equal or the interval between them is finite. The constructibility axiom in T will now well-order the power set of $\omega^*$, creating a metric space composed of the subsets of $\omega^*$. These subsets thus represent the binimals making up the real line $\mathbf{R}^*$.[5] In this theory $\mathbf{R}^*$ is a subset of $\mathbf{Q}^*$.

Now *equality-preserving* bijective mappings between finite intervals of $\mathbf{R}^*$ are homeomorphic, i.e., bijective mappings $\phi(x,u)$ of a finite interval $X$ onto a finite interval $U$ in which $x \in X$ and $u \in U$ such that $\forall x_1, x_2, u_1, u_2 [\phi(x_1, u_1) \wedge \phi(x_2, u_2) \rightarrow (x_1 - x_2 = 0 \leftrightarrow u_1 - u_2 = 0)]$ will produce biunique function pieces which are continuous when taken as either $u(x)$ or $x(u)$. Note that if $X$ or $U$ vanishes then the other must vanish. Finally, the derivatives of these pieces can be obtained by using infinitesimals.

The pieces can be connected so that $\forall x_1, x_2, u_1, u_2 [\phi_1(x_1, u_1) \wedge \phi_2(x_2, u_2) \rightarrow (x_1 - x_2 = 0 \rightarrow u_1 - u_2 = 0)]$ to obtain functions of a real variable $u(x)$: A "function of a real variable $u(x)$ in T" in an interval (a,b) exists if and only if it is a constant (obtained directly from ABR) or a continuously connected sequence of biunique pieces such that the derivative of $u(x)$ with respect to $x$ is also a function of a real variable in T. This statement requires the functions to be continuous, differentiable to all orders and of bounded variation. If some derivative is a constant, they are polynomials. If no derivative is a constant, these functions do not exist in T, as they would require an infinite series description [6]. They can, however, be approached arbitrarily closely by some linear combination of polynomials of sufficiently high degree. These polynomials can be given by multiple iterations of the usual variational form for the Sturm-Liouville problem, so that the results, while still polynomials, become effectively eigenfunctions:

$$\int_a^b [p\left(\frac{du}{dx}\right)^2 - qu^2]dx \equiv \lambda \int_a^b ru^2 dx; \ \lambda \text{ is locally minimum for } \int_a^b ru^2 dx \text{ constant}; \quad (1)$$

where $a \neq b$, $u\left(\dfrac{du}{dx}\right) \equiv 0$ at a and b ; $p, q, r$ are functions of the real variable $x$.

Any algorithm for progressively minimizing $\lambda$ generates increasingly higher degree polynomials. Letting *n* denote the n[th] iteration, we obtain $\lambda_n < \lambda_{n-1}$ such that $\forall k \exists n\ [\lambda_{n-1} - \lambda_n < 1/k]$. By sufficiently high degree we mean empirically indistinguishable when applied to physics (e.g., $k > 10^{50}$).

We show this theory provides a foundation for physical fields governed by a nonlinear sigma model. Let us first consider two eigenfunctions, $u_1(x_1)$ and $u_2(x_2)$; for each let $p \equiv 1$, $q \equiv 0$ and $r \equiv 1$ and we shall call $x_1$ "space" and $x_2$ "time". It is well known that $\left(\frac{\partial u_1 u_2}{\partial x_1}\right)^2 - a\left(\frac{\partial u_1 u_2}{\partial x_2}\right)^2$ is then the Lagrange density for a one-dimensional string $u_1 u_2$ and, by minimizing the integral of this function over all space and time, i.e., by Hamilton's principle, we can determine the physical field arbitrarily closely. We immediately generalize to the Lagrange density for some separable fields in finitely many space-like (i) and time-like (j) dimensions. Since they are functions of real variables in T, the fields so obtained, or any finite sum of such fields, are locally homeomorphic, differentiable to all orders, of bounded variation and therefore without singularities.

Let $u_{\ell m i}(x_i)$ and $u_{\ell m j}(x_j)$ be eigenfunctions with non-negative eigenvalues $\lambda_{\ell m i}$ and $\lambda_{\ell m j}$ respectively. We assert a "field" is a sum of eigenstates: $\underline{\Psi}_m = \sum_\ell \Psi_{\ell m} \underline{i}_\ell$, $\Psi_{\ell m} = \prod_i u_{\ell m i} \prod_j u_{\ell m j}$, subject to the postulate that for every eigenstate *m* the value of the integral of the Lagrange density is *identically* null over space-time. This postulate contains Hamilton's principle. Using $ds d\tau$ for $\prod_i r_i dx_i \prod_j r_j dx_j$,

$$\int \sum_{\ell i} \frac{1}{r_i} \left[ P_{\ell m i} \left(\frac{\partial \Psi_{\ell m}}{\partial x_i}\right)^2 - Q_{\ell m i} \Psi_{\ell m}^2 \right] ds d\tau - \int \sum_{\ell j} \frac{1}{r_j} \left[ P_{\ell m j} \left(\frac{\partial \Psi_{\ell m}}{\partial x_j}\right)^2 - Q_{\ell m j} \Psi_{\ell m}^2 \right] ds d\tau \equiv 0 \quad \text{for all } m. \quad (2)$$

In this integral expression the *P* and *Q* can be functions of any of the $x_i$ and $x_j$, thus of any $\Psi_{\ell m}$ as well. This is a *nonlinear sigma model*. The $\Psi_{\ell m}$ can be determined by an algorithm with multiple iterations that are constrained by the indicial expression $\sum_{\ell i} \lambda_{\ell m i} - \sum_{\ell j} \lambda_{\ell m j} \equiv 0$ for all *m*. [7]

We now show that expression (2) *requires quantization*. We will denote each of the integral terms individually by $\alpha$, since they are identical. Here is a proof in T that $\alpha$ has only discrete finite values:

I      $\alpha$ is positive and must be closed to addition and to the absolute value of subtraction;
        In T we must have that $\alpha$ is an integer times a constant which is either infinitesimal or finite.

II     There is either no field (and no space-time), in which case $\alpha \equiv 0$,
        or otherwise in T the field is finite, in which case $\alpha \neq 0$; thus $\alpha = 0 \leftrightarrow \alpha \equiv 0$.

III    $\therefore \alpha \equiv nI$, where $n$ is an integer and $I$ is a finite constant such that $\alpha = 0 \leftrightarrow n \equiv 0$.

Quantization of the field is thus obtained from a physical null postulate added to the mathematics of T, in which the definition of functions of a real variable requires space-time to vanish if the field vanishes.

When there are finitely many space-like dimensions with just one time-like dimension, we can obtain the Schrödinger equation from expression (2): Let $\ell = 1,2$ and suppress $m$. We introduce:

$\Psi = A \prod_i u_i(x_i)[u_1(\tau) + \iota u_2(\tau)]$, where $\iota = \sqrt{-1}$, normalizing $[u_1^2(\tau) + u_2^2(\tau)] \int \prod_i u_i^2(x_i)\, ds \equiv 1$. Then we see that $\dfrac{du_1}{d\tau} = -u_2$ and $\dfrac{du_2}{d\tau} = u_1$ or $\dfrac{du_1}{d\tau} = u_2$ and $\dfrac{du_2}{d\tau} = -u_1$. In either case, for each and every *irreducible* biunique time-eigenfunction piece the value of $\alpha$ must be *least*, i.e., just the finite constant $I$.

Thus $A^2 \int \int_0^{\pi/2} \left[\left(\dfrac{du_1}{d\tau}\right)^2 + \left(\dfrac{du_2}{d\tau}\right)^2\right] \prod_i u_i^2(x_i)\, ds\, d\tau \equiv A^2 \pi/2 \equiv I$. Substituting the Planck constant h for $4I$ and letting $\tau$ be $\omega t$, the integrand becomes the well-known time-part of the Lagrange density for the Schrödinger equation, $(h/4\pi\iota)\left[\Psi^*\left(\dfrac{\partial \Psi}{\partial t}\right) - \left(\dfrac{\partial \Psi^*}{\partial t}\right)\Psi\right]$. The energy in the $m^{th}$ eigenstate, defined here as $A^2 \dfrac{d\tau_m}{dt}$, where $\tau_m = \omega_m t$, exists only in quanta of $h\omega_m/2\pi$. The sum of the energies in all of the eigenstates $E_t$ is thus $\sum n_m h\omega_m/2\pi$ where $n_m$ is the number of occupants of the $m^{th}$ eigenstate.

We can now finish the derivation of the definability in T of any field obtained from expression (2) in finitely many space-like dimensions and one time dimension in a finite region of space. By the fundamental theorem of arithmetic, every ordered set of $n_m$, representing a unique distribution of an energy $E_t$ among the eigenstates of the field, corresponds to a unique finite integer and every finite integer corresponds to a unique set of $n_m$, as follows:

$$\{n_m\} \Leftrightarrow \prod_m [P_m]^{n_m} \text{ where } P_m \text{ is the } m^{th} \text{ prime starting with 2.} \quad (3)$$

A set (in T) of these finite integers exists for all finite $E_t$. So quantization implies definability for a finite $E_t$. Moreover, in T we also can show the converse, that definability implies quantization. If "=" appears in expression (2) instead of "≡", then the requirement for quantization no longer holds and the set of all possible distributions of energy among the eigenstates cannot be mirrored by a set (in T) of finite integers. Thus *definability in T is equivalent to quantization*. By similar reasoning, definability in T can be shown equivalent to compactification of all the spatial dimensions effectively [8].

In addition to providing a foundation for quantized physical fields, here are three examples of the applicability of theory T to physics. First, in T fields are basically constructed from smoothly connected biunique eigenfunction pieces. As we have seen, these pieces result from a homeomorphic mapping which is symmetric between range and domain. The construction necessitates not only that there are no discontinuities of the field but also if the field vanishes space-time vanishes, i.e., space-time is relational. Moreover, all singularities that appear at the Fermi scale must be resolved at the Planck scale, since fields, as given by expression (2), can have no singularities. Second, the well-known argument of Dyson [9], that the perturbation series used in quantum electrodynamics are divergent (thus must be merely asymptotic expansions that in practice give an accurate approximation), is avoided in this theory, i.e., any series in this theory must be finite, so perturbations are limited to some finite order, however large, and his proof fails. These examples suggest that theory T may offer a possible foundation for quantum gravity. Third, the deep question raised by Wigner [10] about the unreasonable effectiveness of mathematics in physics is here answered directly.

Acknowledgement: The author's thanks to Jan Mycielski of the University of Colorado for confirming the consistency of the theory, as well as to Vatche Sahakian, Maksim Perelstein and Kurt Gottfried of Cornell University and Rathin Sen of Ben-Gurion University for their advice and comments.

1. Tarski, A., Mostowski, A., Robinson, R., *Undecidable Theories*. North Holland, Amsterdam, 1953.
2. Gödel, K., The consistency of the axiom of choice and of the generalized continuum hypothesis. *Annals of Math. Studies*, 1940.
3. Cohen, P. J., *Set Theory and the Continuum Hypothesis*, New York, 1966.
4. The axiom schema of subsets is $\exists u[[u = 0 \vee \exists x x \in u] \wedge \forall x x \in u \leftrightarrow x \in z \wedge X(x)]$, where $z$ is any set and $X(x)$ is any formula in which $x$ is free and $u$ is not free. The axiom enters ZF in AR but can also enter in the strong form of the axiom of regularity. (Note T has the weak form.) This axiom is essential to obtain the diagonal set for Cantor's proof, using $x \notin f(x)$ for $X(x)$, where $f(x)$ is an assumed one-to-one mapping between $\omega^*$ and $P(\omega^*)$. The argument leads to the contradiction $\exists c \in z X(c) \leftrightarrow \neg X(c)$, where $f(c)$ is the diagonal set. In ZF, this denies the mapping exists. In T, the same argument instead denies the existence of the diagonal set, whose existence has been hypothesized while the mapping was asserted as an axiom. What if we tried the approach of using ABR to get a characteristic function? Let $\phi(x,y) \leftrightarrow [X(x) \leftrightarrow y = (x,1) \wedge \neg X(x) \leftrightarrow y = (x,0)]$ and $z = \omega^*$. If c were a member of $\omega^*$, $t = (c,1)$ and $t = (c,0)$ both lead to a contradiction. But, since the existence of the diagonal set $f(c)$ is denied and since a one-to-one mapping between $\omega^*$ and $P(\omega^*)$ is an axiom, as $f(c)$ is not a member of $P(\omega^*)$, so c cannot be a member of $\omega^*$. In T the characteristic function exists but has no member corresponding to a diagonal set.
5. The axiom of constructibility generates sequentially all the subsets of $\omega^*$ in a set of ordered pairs. The left-hand member of each pair is a subset of $\omega^*$ and the right-hand member is an integer indicating the order in which it was generated. If we let the integers not present in each subset be a "1" in the corresponding binimal and the integers that are present be a "0", then the right-hand member is the magnitude of that binimal and serves as a distance measure on the line $R^*$.
6. We have required the physical universe to be logical and this logic to be two-valued. Therefore physical behavior will be digital, that is, the universe becomes fundamentally a logic machine or "computer". Infinite series cannot be calculated *exactly* on a computer since we cannot have an infinite number of iterations.
7. The $u_{\ell m i}(x_i)$ and $u_{\ell m j}(x_j)$ are iterated using (1). The $p_{\ell m i}(x_i), q_{\ell m i}(x_i), p_{\ell m j}(x_j)$ and $q_{\ell m j}(x_j)$ will generally change at each iteration and are given by $p_{\ell m i} = \int \frac{P_{\ell m i} \Psi_{\ell m}^2 d\tau}{u_{\ell m i}^2 r_i dx_i} \bigg/ \int \frac{\Psi_{\ell m}^2 d\tau}{u_{\ell m i}^2 r_i dx_i}$ , etc.

    Since the field is continuous, differentiable to all orders and of bounded variation, iterations for all $u_{\ell m i}(x_i)$ and $u_{\ell m j}(x_j)$, given the constraint $\sum_{\ell i} \lambda_{\ell m i} - \sum_{\ell j} \lambda_{\ell m j} \equiv 0$, will converge jointly within a finite region.
8. The same reasoning can be applied to the spatial dimensions. A field is definable in T if and only if all m are finite. In T, the range and domain of the irreducible biunique eigenfunction pieces in each of the spatial dimensions is finite (i.e., is not infinitesimal or infinite) and all functions are continuous. Therefore, if any spatial dimension is infinite, some m are infinite and the field is not definable. If all spatial dimensions are finite, we have shown that the field is quantized, hence definable in T. The field is thus definable in T if and only if all the spatial dimensions are finite. So compactification of the spatial dimensions is equivalent to quantization. We have obtained compactification effectively and this was achieved without invoking any particular boundary conditions.
9. Dyson, F. J., Divergence of Perturbation Theory in Quantum Electrodynamics, Phys.Rev., 1952, 85.
10. Wigner, E. P., The Unreasonable Effectiveness of Mathematics in the Natural Sciences, Comm. Pure and Appl. Math. 1960, 13.

# Appendix

## ZF - AR + ABR + Constructibility

Extensionality   Two sets with just the same members are equal.

-   $$\forall x \forall y [\forall z [z \in x \leftrightarrow z \in y] \to x = y]$$

Pairs -   For every two sets, there is a set that contains just them.

$$\forall x \forall y \exists z [\forall w\, w \in z \leftrightarrow w = x \vee w = y]$$

Union -   For every set of sets, there is a set with just all their members.

$$\forall x \exists y \forall z [z \in y \leftrightarrow \exists u [z \in u \wedge u \in x]]$$

Infinity -   A set $\omega^*$ contains 0 and each member has a next member containing just all its predecessors.

$$\exists \omega^* [0 \in \omega^* \wedge \forall x [x \in \omega^* \to x \cup \{x\} \in \omega^*]]$$

Power Set -   For every set, there is a set containing just all its subsets.

$$\forall x \exists P(x) \forall z [z \in P(x) \leftrightarrow z \subseteq x]$$

Regularity -   Every non-empty set has a minimal member (i.e. "weak" regularity).

$$\forall x [\exists y\, y \in x \to \exists y [y \in x \wedge \forall z \neg [z \in x \wedge z \in y]]]$$

Replacement -   Replacing members of a set one-for-one creates a set (i.e., "bijective" replacement).

Let $\phi(x,y)$ a formula in which x and y are free,

$$\forall z \forall x \in z \exists y [\phi(x,y) \wedge \forall u \in z \forall v [\phi(u,v) \to u = x \leftrightarrow y = v]] \to \exists r \forall t [t \in r \leftrightarrow \exists s \in z\, \phi(s,t)]$$

Constructibility -   All the subsets of any $\omega^*$ are constructible.

$$\forall \omega^* \exists S [(\omega^*, 0) \in S \wedge \forall y E! z [y \neq 0 \wedge y \subseteq \omega^* \wedge (y,z) \in S \leftrightarrow (y \cup m_y - \{m_y\}, z \cup \{z\}) \in S]]$$

where $m_y$ is the minimal member of y.